\documentclass[12pt]{article}
\usepackage{amsmath}
\usepackage{amssymb}
\usepackage{graphicx}
\usepackage{dsfont}
\usepackage{enumerate}
\usepackage{natbib}
\usepackage{booktabs}
\usepackage{array}
\usepackage{siunitx}
\usepackage{textcmds}
\newcommand{\blind}{1}
\usepackage[font=footnotesize,labelfont=bf]{caption}
\usepackage[ruled]{algorithm2e}
\usepackage{array}
\usepackage{bbm}
\usepackage{multirow}
\usepackage[parfill]{parskip}
\usepackage[utf8]{inputenc}
\usepackage{amsfonts}
\usepackage{enumitem}
\usepackage{amsmath}
\usepackage{soul}
\usepackage{natbib}

\usepackage{fancyhdr}
\usepackage{float}
\usepackage{setspace}

\usepackage{titling}
\usepackage{graphicx}
\usepackage[margin=1in]{geometry}
\usepackage{csquotes}
\usepackage{amssymb}
\usepackage{mathrsfs}
\usepackage[hidelinks]{hyperref}

\def\bSig\mathbf{\Sigma}
\usepackage[ruled]{algorithm2e}\usepackage[font=footnotesize,labelfont=bf]{caption} 
\usepackage{xcolor}
\usepackage[margin=1in]{geometry}

\begin{document}

\def\spacingset#1{\renewcommand{\baselinestretch}%
{#1}\small\normalsize} \spacingset{1}

\if1\blind
{
  \title{\bf Federated Survival Analysis with Site-Level Heterogeneity Adjustment}
  \author{
  Hyojung Jang\\
  Division of Biostatistics, Department of Preventive Medicine, \\ Northwestern University \\
  \\
  Malcolm Risk\\
    Department of Biostatistics, University of Michigan\\
    \\ 
    Yaojie Wang\\
    Department of Preventive Medicine, Northwestern University \\
    \\
  Norrina Bai Allen\\
  Department of Preventive Medicine, \\ Northwestern University \\
  \\
     Xu Shi \\
    Department of Biostatistics, University of Michigan \\
    \\
    Lili Zhao \\
   Division of Biostatistics, Department of Preventive Medicine, \\ Northwestern University }
    \date{}  %
  \maketitle
} \fi

\bigskip
\begin{abstract}
In multi-center clinical research, privacy regulations often prohibit pooling individual-level records, complicating the analysis of time-to-event data. Current federated survival methods frequently require iterative communication or rely strictly on proportional hazards (PH) assumptions or require sensitive survival information. We propose a one-shot federated framework using pseudo-observations derived from a sequentially updated Kaplan-Meier estimator and fitted via a renewable generalized estimating equation. Unlike traditional methods, our approach allows flexible link functions tailored to the target estimand and accommodates non-proportional hazards. To address site-level heterogeneity, we introduce a covariate-wise debiasing procedure that shrinks noise-driven local deviations toward the global estimate while preserving genuine site-specific effects. Simulation studies demonstrate that our framework achieves inferential accuracy comparable to pooled Cox regression and the privacy-preserving One-shot Distributed Algorithm to fit a multicenter Cox proportional hazards model (ODAC) under PH assumptions, while recovering time-varying coefficient trajectories when PH is violated. Furthermore, simulations confirm that the debiasing procedure optimizes the bias-variance trade-off, adaptively balancing global stability with the preservation of genuine site-specific deviations. Applied to pediatric obesity data from the Chicago Area Patient-Centered Outcomes Research Network (CAPriCORN) network ($N=45,865$), the model produced robust estimates of time-invariant and time-varying hazard ratios, offering a flexible, privacy-preserving alternative for collaborative survival research.

\end{abstract}

\newpage
\spacingset{1.9} 
\section{Introduction}
\label{sec:intro}

Time-to-event outcomes—such as time to death, time to disease progression, and time to serious adverse events—are central endpoints in clinical research. Survival analysis focuses on these outcomes by modeling the distribution of time-to-events and quantifying the risk of experiencing an event over follow-up. The widespread adoption of real-world data infrastructures, including electronic health records (EHRs) \citep{sherman2016, concato2022}, insurance claims \citep{ahn2022, baser2023}, and wearable devices \citep{izmailova2018, daniore2024}, has greatly expanded access to longitudinal patient-level information. Unlike cross-sectional datasets, these longitudinal records capture follow-up and event information and provide a foundation for time-to-event modeling in real-world settings. 

The increasing availability of real-world data (RWD) within individual healthcare institutions has created new opportunities for survival research. However, studies conducted at a single center often remain underpowered and lack generalizability, particularly when targeting rare clinical events or narrowly defined patient subgroups that are sparsely represented at one site. As a result, there is a growing need for multi-center survival analyses. Yet, sharing individual-level EHR data across organizations raises substantial concerns related to patient privacy, data governance, and institutional ownership, and is frequently constrained by complex regulatory, contractual, and technical requirements. These barriers make traditional pooled analyses slow, costly, and difficult to implement at scale. Such challenges have motivated increasing interest in federated learning (FL) approaches, which enable cross-institutional collaboration without transferring individual-level records, thereby supporting multi-site time-to-event analyses while preserving local control over sensitive data \citep{vogelsang2020, spath2022, froelicher2021, bonomi2020}.

The Cox proportional hazards model \citep{Cox1972} remains the standard for analyzing time-to-event outcomes for assessing associations between covariates and time-to-event outcomes. Several privacy-preserving approaches have been proposed for multicenter survival analysis. Early work includes secure multi-party computation and homomorphic encrpytioni-based methods, such as Lu et al \citep{lu2021}, and iterative federated algorithms, such as WebDISCO \citep{Lu2015}, which approximate pooled estimates through multiple rounds of summary statistics exchange. However, both approaches are primarily focused on point estimation and are limited in their ability to perform full statistical inference; furthermore, cryptographic methods like Lu et al. \citep{lu2021} often impose heavy computational burdens. To improve efficiency, non-iterative non-iterative \qq{one-shot} methods have emerged, including ODAC \citep{Duan2020} and risk-set based approaches \citep{li2023} have emerged, constructing surrogate likelihoods or estimating equations to approximate global Cox model, requiring only a single round of transmission of aggregate quantities. Duan further extended this framework with ODACH \citep{luo2022odach} to accommodate site heterogeneity.

Among these approaches, ODAC \citep{Duan2020} serves as a representative one-shot benchmark. ODAC approximates the pooled Cox estimator by constructing a surrogate partial likelihood at each site. The process begins with an initial exchange of local coefficient estimates and distinct event times to establish a global initialization. Using this shared baseline, sites compute local gradient and Hessian matrices to maximize their respective surrogate likelihoods, which can be subsequently aggregated via inverse-variance weighting. This framework achieves a communication-efficient approximation of the pooled model without requiring iterative data transfer. However, the algorithm requires participating sites to share the set of unique event times across the network, which may still reveal sensitive information about the underlying survival process.

Despite their practical advantages, existing federated Cox procedures share important limitations. Their reliance on the proportional hazards assumption restricts flexibility in handling time-varying effects and limits the direct estimation of survival probabilities. Furthermore, sharing sensitive survival information such as event times and number at risks raises residual privacy concerns, while cryptographic alternatives impose heavy computational burdens. These considerations motivate the development of federated methods that move beyond Cox PH models, directly target survival probabilities, accommodate both time-invariant and time-varying effects, and incorporate mechanisms to account for site-level heterogeneity. To address these needs, we introduce a federated survival modeling framework based on federated pseudo-observations. Pseudo-observations are constructed from a sequentially updated federated Kaplan–Meier estimator \citep{malcolm2025} and analyzed using renewable generalized linear models (GLMs) \citep{luo2020}, treating the pseudo-values as outcomes.

This methodology advances federated survival analysis through two primary contributions. First, we develop a flexible federated pseudo-value regression framework that does not rely on the proportional hazards assumption. The method is formulated within a generalized estimating equation (GEE) framework, which allows different link functions to be specified according to the target estimand. For example, the logit link enables direct estimation of odds ratios within the same modeling framework. The complementary log–log link further allows both proportional and non-proportional hazards to be accommodated through time-invariant or time-varying effects. To ensure valid statistical inference, we embed a sandwich variance estimator within the renewable GLM, providing reliable variance estimates in the presence of repeated measurements and clustered observations.

Second, we introduce a debiasing procedure to capture heterogeneity in covariate effects across sites. In federated settings, some centers may exhibit site-specific effects that differ from the overall population pattern, whereas most existing multicenter survival models assume common regression coefficients across sites. In contrast, we explicitly model heterogeneity in covariate effects, allowing regression coefficients to vary across sites when supported by the data. To achieve this, we adopt a “fit-and-adjust” strategy: after fitting the global federated pseudo-value regression, each site applies a covariate-wise soft-thresholding adjustment \citep{donoho1994} to its local coefficient vector, shrinking estimates toward the global target. The amount of shrinkage is determined using a Generalized Stein’s Unbiased Risk Estimate (GSURE) \citep{xie2012sure}, which provides a data-driven rule for selecting the threshold while accounting for the variance and correlation structure of the site-specific deviations. This design is tailored for sparse heterogeneity, preserving genuine local signals while maintaining efficiency through information sharing. Together, these components provide a flexible and privacy-preserving framework for multicenter time-to-event analysis that achieves inference comparable to pooled analyses while capturing complex survival dynamics.

\section{Methods}
\label{sec:meth}

\subsection{Notation}

We begin by defining key notations in survival analysis. Let $T_i$ and $C_i$ denote the true event time and censoring time, respectively, for subject $i = 1, \dots, N$. We observe the follow-up time $X_i = \min(T_i, C_i)$ and the event indicator $\Delta_i = \mathbb{I}(T_i \le C_i)$, where $\mathbb{I}(\cdot)$ is the indicator function. Let $Z_i$ denote a $p$-dimensional vector of baseline covariates for subject $i$. The observed data for the cohort is denoted by $\mathcal{D} = \{(X_i, \Delta_i, Z_i)\}_{i=1}^N$.

When we consider a federated network comprising $K$ clinical sites. The data are partitioned such that the $k$-th site holds $n_k$ subjects, yielding a total sample size of $N = \sum_{k=1}^K n_k$. At site $k$, the observed data for subject $i$ are denoted by $\mathcal{D}_k = \{(X_{ik}, \Delta_{ik}, Z_{ik})\}_{i=1}^{n_k}$.

\subsection{Review Pseudo-value Regression in a Single Site}

In centralized settings where individual-level data $(X_i,\Delta_i,Z_i)$ are fully available, the marginal survival function $S(t)=P(T>t)$ can be estimated using the Kaplan--Meier (KM) estimator
\begin{equation}
\hat S(t)=\prod_{t_k\le t}\left(1-\frac{d_k}{Y_k}\right),
\end{equation}
where $t_1<t_2<\cdots$ denote the ordered distinct event times, $d_k$ is the number of events at time $t_k$, and $Y_k$ is the number of individuals at risk immediately prior to $t_k$. Based on this estimator, the pseudo-observation \citep{andersen2010} for subject $i$ evaluated at a prespecified landmark time $t_j$ is constructed via a leave-one-out (jackknife) approach, which is  defined as  $\tilde S_{ij}=N\hat S(t_j)-(N-1)\hat S^{(-i)}(t_j),$ where $\hat S(t_j)$ is the KM estimator computed using all $N$ subjects and $\hat S^{(-i)}(t_j)$ is the KM estimator computed after removing subject $i$ from the sample. These pseudo-observation $\tilde S_{ij}$ are treated as continuous outcomes and modeled using a generalized linear model (GLM) to estimate covariate effects $Z_i$ on the survival probabilities at time $t_j$ :
\begin{equation}
g\!\left(E[\tilde S_{ij}\mid Z_i]\right)=\beta_0+Z_i^\top\beta ,
\end{equation}
where $g(\cdot)$ denotes a link function. The interpretation of $\beta$ depends on the choice of link function. For example, the complementary log-log (cloglog) link $g(x)=\log(-\log x)$ yields regression coefficients that correspond to log hazard ratios (HRs), whereas a logit link produces a proportional odds interpretation.

To model survival probabilities over the follow-up period, pseudo-observations are typically evaluated at multiple landmark times $t_1,\dots,t_J$ \citep{andersen2010,anyaso2023,zhao2020deep,zhao2020deepBioinformatics}. It has been shown that as few as five time intervals equally spread on the event time scale worked quite well in most cases \citep{zhao2020deep,zhao2020deepBioinformatics}. Stacking these pseudo-observations yields repeated outcomes for each subject. Let $\theta_i=(\tilde S_{i1},\dots,\tilde S_{iJ})^\top$ denote the vector of pseudo-observations for subject $i$. Regression parameters are estimated by solving the generalized estimating equations (GEE):
\begin{equation}
\sum_i U_i(\beta)= \sum_{i}
\left(\frac{\partial}{\partial\beta} g^{-1}(\beta^\top Z_i^*)\right)^\top
V_i^{-1}
\left(\theta_i-g^{-1}(\beta^\top Z_i^*)\right)=0,
\end{equation}
where $Z_i^*$ denotes the expanded design matrix including covariates and landmark time indicators, and $V_i$ is a working covariance matrix describing the within-subject dependence among the pseudo-observations. Because each subject contributes multiple pseudo-observations, inference is obtained using a robust sandwich variance estimator to account for the induced within-subject correlation.

Under the cloglog link and the constraint $\beta(t_j)=\beta$, the resulting estimates correspond to log HRs under the Cox proportional hazards model. Allowing $\beta(t_j)$ to vary across landmark times enables modeling of time-varying covariate effects using either the cloglog link for hazards or the logit link for odds.

\subsection{Pseudo-value Regression in Federated Settings}

We consider a federated network comprising $K$ clinical sites where the privacy constraints prohibit the central pooling of individual-level data. To perform pseudo-value regression in this setting, we adopt a two-stage federated framework.

First, pseudo-observations are constructed locally using the global KM estimator $\hat S(t)$ \citep{malcolm2025} together with the influence function  $\hat\psi(t)$. The pseudo-value for subject $i$ at landmark time $t_j$ is approximated as
\begin{equation}
\tilde S_{ij} \approx \hat S(t_j) + \hat\psi_i(X_i, \Delta_i),
\label{eq:pseudo}
\end{equation}

where $\hat\psi_i(t_j)$ denotes the empirical influence function evaluated at $i^{th}$ patient data $(X_i, \Delta_i)$ at time $t_j$. Specifically,
\begin{equation}
\hat \psi(X_i, \Delta_i)(t) =
-
\hat S(t)
\left[
\frac{\Delta_i \mathbb{I}(X_i \le t)}{\hat Y(X_i)}
-
\int_0^{X_i \wedge t}
\frac{\hat \lambda(u)}{\hat Y(u)}du
\right],
\label{eq:IF}
\end{equation}

where $Y(u)=P(X>u)$ denotes the probability of remaining at risk at time $u$ and $\lambda(u)$ denotes the hazard function. 

Pseudo observation construction using Equation (\ref{eq:pseudo}) provides a first-order approximation to the classical leave-one-out pseudo-value construction, avoiding the computational burden of repeatedly recomputing the KM estimator after removing each subject. This is particularly important for large datasets, where the leave-one-out estimator is computationally infeasible. In the federated setting, each site receives the global estimators $\hat S(t)$ and $\hat\psi(t)$ and can then evaluate $\tilde S_{ij}$ locally using these quantities together with its own patient-level data, thereby preserving privacy.

Second, regression coefficients are estimated using a renewable generalized estimating equation (GEE) framework, where the pseudo-observations constructed in the first step are outcomes. This approach sequentially updates the estimating equations across sites without pooling individual-level data. Let $\tilde{\mathbf S}_i = (\tilde S_{i1},\dots,\tilde S_{iJ})^\top$ denote the vector of pseudo-observations for subject $i$ evaluated at landmark times $t_1,\dots,t_J$. The evaluation times are chosen from quantiles of the KM survival curve, corresponding to time points approximately evenly spaced on the event scale. The regression parameters $\beta$ are estimated by solving the stacked estimating equations

\begin{equation}
\sum_{k=1}^{K} U_k(\beta) = 0,
\end{equation}

where $U_k(\beta)$ denotes the score contribution computed from site $k$. In the renewable estimation procedure, the algorithm initializes at Site 1 to obtain the local estimate $\hat\beta^{(1)}$. For $k=2,\dots,K$, Site $k$ updates the parameter estimate by solving:$$\tilde H_{k-1}(\hat\beta^{(k-1)}-\beta) + U_k(\beta) = 0,$$relying on the previous estimate $\hat\beta^{(k-1)}$ and the cumulative negative Hessian $\tilde H_{k-1} = \sum_{\ell=1}^{k-1} H_\ell(\hat\beta^{(\ell)})$ transmitted from preceding sites. This equation provides a first-order approximation to the pooled estimating equation and can be solved using standard Newton–Raphson updates \citep{luo2020}.

Since each subject has $J$ pseudo-observations at $J$ time points, we propose using  a robust sandwich variance estimator in the above renewable framework, extended to account for within-subject correlation. This requires transmitting the cumulative negative Hessian ($\tilde{H}_k$) and the cumulative meat matrix: $\tilde{M}_k = \sum_{\ell=1}^k \sum_{i=1}^{n_\ell} U_\ell^{(i)}\bigl(\hat\beta^{(\ell)}\bigr) U_\ell^{(i)}\bigl(\hat\beta^{(\ell)}\bigr)^\top$, where $U_\ell^{(i)}\bigl(\hat\beta^{(\ell)}\bigr)$ is the $q \times 1$ subject-level score vector for subject $i$ at site $\ell$, and $q$ denotes the total number of parameters in the model, including baseline covariates, time indicators, and their interactions when modeling time-varying effects. At the final site $K$, the global robust variance estimator is given by:$$\widehat V(\hat\beta) = \tilde H_K^{-1} \tilde M_K \tilde H_K^{-1}.$$

The complete federated estimation procedure, including pseudo-observation construction and renewable GEE updates, is summarized in Algorithm~\ref{algorithm:federated_pseudo}.

\spacingset{1.0}
\begin{algorithm*}[!bp]
\caption{\label{algorithm:federated_pseudo} Federated Pseudo-value Regression for Proportional Hazards Survival Models}
\small

\textbf{Inputs:} Data $\{(X_{ik},\Delta_{ik},Z_{ik}): i=1,\dots,n_k;\ k=1,\dots,K\}$ stored locally;  $N=\sum_{k=1}^K n_k$.

\textbf{Outputs:}  Global estimates of  $\hat{\beta}$ and sandwich variance $\widehat V(\hat{\beta})$ using data from all $K$ sites in privacy-preserving way

\medskip
\textbf{Algorithm:}

\medskip
\noindent
\textbf{Step 1: Pseudo-observation construction in federated way}

\begin{itemize}

\item Construct the global Kaplan–Meier function $\hat S(t)$ and corresponding Influence function $\hat \phi(t)$ and broadcast them to each local site.

\item At each site $k$, compute pseudo-observations for subject $i$ at $t_j$ ($j=1,\dots,J$)
\[
\tilde S_{ij} = \hat S(t_j) + \hat \psi_i (X_i, \Delta_i) (t_j),
\]

\end{itemize}

\medskip
\noindent
\textbf{Step 2: Renewable GEE estimation}

\begin{itemize}

\item Fit a GEE using the pseudo-observations at site 1 to obtain the local estimate $\hat\beta^{(1)}$.

\item For $k = 2,\dots,K$, update the estimate $\hat\beta^{(k)}$ sequentially by solving the renewable estimating equation
\[
\tilde{H}_{k-1}(\hat{\beta}^{(k-1)} - \beta) + U_k(\beta) = 0.
\]

\item Update the cumulative summary matrices over site. 
\[
\tilde{H}_k = \tilde{H}_{k-1} + H_k(\hat{\beta}^{(k)}),
\qquad
\tilde{M}_k = \tilde{M}_{k-1} + M_k.
\]
\item Return global estimates: $\hat\beta = \hat\beta^{(K)}$ and  $\widehat V(\hat\beta) = \tilde H_K^{-1} \tilde M_K \tilde H_K^{-1}.$
\end{itemize}

\medskip
\noindent

\end{algorithm*}
\spacingset{1.9}
\medskip
\noindent

\subsection{Addressing Site Heterogeneity}
In federated networks, regression coefficients may vary across sites due to differences in patient populations, clinical practice, or unmeasured confounding. To accommodate such heterogeneity, we adopt a fit-and-adjust strategy that shrinks site-specific estimates toward the  global estimate while allowing meaningful deviations to persist.

To accommodate site heterogeneity, we consider the site-specific deviation
$\Delta_k=\hat{\beta}^{(k)}-\hat{\beta}_{\mathrm{glob}}$, $k=1,\ldots,K$.
We then apply variance-adaptive soft-thresholding to shrink deviations that are small
relative to their uncertainty while retaining larger departures from the global estimate:
\[
\hat{\delta}_k(\tau)
=
\operatorname{sign}(\Delta_k)\left(|\Delta_k|-\tau\,\sqrt{V(\Delta_k)}\right)_+,
\]
where $(x)_+=\max(x,0)$. Here, $\tau$ is a global tuning parameter and the effective threshold for site $k$ is 
$\tau\sqrt{V(\Delta_k)}$, where $V(\Delta_k)$ represents the variance of the deviation 
between the local and global estimators, that is, $V(\Delta_k) = \mathrm{Var}(\hat{\boldsymbol{\beta}}^{(k)}) + \mathrm{Var}(\hat{\boldsymbol{\beta}}_{\mathrm{glob}}) - 2\mathrm{Cov}(\hat{\boldsymbol{\beta}}^{(k)}, \hat{\boldsymbol{\beta}}_{\mathrm{glob}}).$  We estimate $\mathrm{Cov}(\hat{\boldsymbol{\beta}}^{(k)}, \hat{\boldsymbol{\beta}}_{\mathrm{glob}})$ using components of the robust sandwich variance estimator, approximated by $H^{-1} M_k H_k^{-\top}$ where $H$ and $H_k$ denote the global and site-specific negative Hessian matrices, respectively, and $M_k$ is the local contribution to the sandwich \qq{meat}, constructed from the outer product of the site-level score contributions.
Under this construction, sites with larger uncertainty in the deviation are shrunk more strongly toward the global estimate. 

Because the threshold is scaled by $V(\Delta_k)$, sites with greater uncertainty in their deviation estimates receive stronger shrinkage toward the global estimate. In particular, $V(\Delta_k)$ increases when the local estimator has larger variance or weaker correlation with the global estimator, leading to stronger shrinkage. Conversely, when the local estimator is highly correlated with the global estimator, the local estimate closely tracks the global estimate, reducing the need for additional shrinkage.

The shrinkage parameter $\tau$ is selected using a Generalized Stein's
Unbiased Risk Estimate (GSURE) \citep{eldar2008generalized}. GSURE is applied
because the deviations $\Delta=(\Delta_1,\ldots,\Delta_K)$ are correlated,
violating the independent Gaussian sequence assumption underlying the
classical SURE framework. Under this setting, the GSURE criterion used for
selecting $\tau$ is
\[
\mathrm{GSURE}(\tau)
=
\sum_{k=1}^K
\left[
\hat{\delta}_k(\tau)^2
-
2\hat{\delta}_k(\tau)\Delta_k
+
2V(\Delta_k)\mathbb{I}\!\left(|\Delta_k|>\tau\sqrt{V(\Delta_k)}\right)
\right].
\]

The optimal threshold $\hat{\tau}$ is obtained by minimizing this criterion
over a finite grid.

Finally, the adjusted  estimate for site $k$ is constructed as
\[
\hat{\boldsymbol{\beta}}_{\mathrm{deb}}^{(k)}
=
\hat{\boldsymbol{\beta}}_{\mathrm{glob}}
+
\hat{\delta}_k(\hat{\tau}), \mbox{~~~} k=1,\cdots,K
\]

The GSURE framework provides the flexibility to accommodate non-i.i.d. data in federated settings. To ensure numerical stability when the number of sites ($K$) is limited, we shrink local estimates toward the global estimator as a prespecified target, rather than pursuing the simultaneous estimation of a data-driven shrinkage location. Following the methodology of \citet{xie2012sure}, we utilize the global estimator as a predetermined grand-mean anchor to mitigate overfitting and preserve asymptotic optimality in federated clinical environments where $K$ is typically modest. Notably, while this target is functionally assigned to stabilize the estimation, our method recognizes its stochastic nature; unlike approaches that treat shrinkage targets as independent constants \citep{huang2025}, the proposed rule explicitly accounts for the covariance between local and global estimators, yielding a correlation-aware and variance-adaptive shrinkage mechanism.

\section{Simulation Studies}

We conducted simulation studies to assess the proposed framework's performance under two  settings when data is generated under the PH assumption and  under the non-PH assumption. Additionally, we evaluated the SURE-based soft-thresholding debiasing procedure under sparse site heterogeneity to demonstrate its optimal bias-variance trade-off in recovering local effects.

\subsection{Scenario 1: Data Generated under the Proportional Hazard Assumption}

We evaluated the performance of the federated pseudo-value regression—fitted via a renewable GEE with a complementary log-log link—against the oracle pooled Cox model and ODAC \citep{Duan2020}. Data were generated with a total sample size of $N = 7{,}500$ under four configurations defined by two event rates, chosen to represent rare (10\%) and moderately common (30\%) outcomes, and two site-network structures: (1) $K = 5$ sites, and (2) $K = 10$ sites. Within each network structure, we varied the per-site sample sizes while maintaining the total sample size.

We generated a continuous confounder $X \sim \mathcal{N}(0, 1)$ and a binary treatment indicator $A \sim \text{Bernoulli}(\text{expit}(0.5X))$. Survival times followed an exponential distribution with hazard $\lambda = h_0 \exp(\beta_X X + \beta_A A)$, where the true coefficients were set to $\beta_X = \log(0.7)$ and $\beta_A = \log(1.15)$.  The constant baseline hazard $h_0$ was calibrated to achieve the target event rates under administrative censoring. 

\begin{figure}[htbp]
  \centering
  \includegraphics[width=0.9\textwidth]{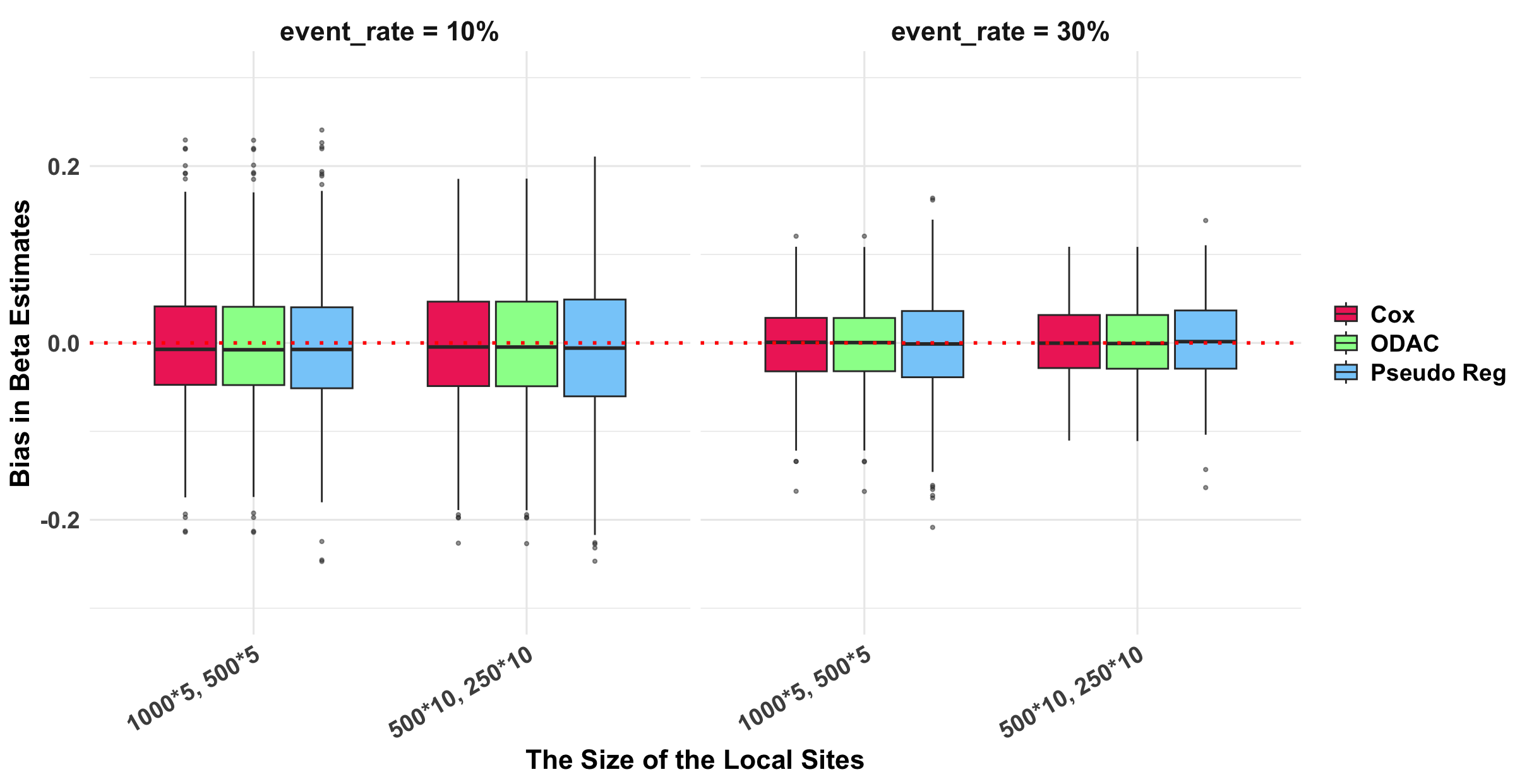} 
  \caption{Boxplots of bias in the estimated treatment log-HR across 500 simulation replicates under the PH setting ($N = 7{,}500$). The left and right panels correspond to 10\% and 30\% event rates, respectively. The x-axis denotes the site configuration (e.g., ``$1000 \times 5$'' indicates 5 sites with 1,000 subjects per site). The pooled Cox model serves as the reference.}
  \label{fig:sim_ph}
\end{figure}

Figure~\ref{fig:sim_ph} presents the bias of the estimated treatment log-hazard ratios (log-HRs) across 500 simulation replicates. Using the Cox proportional hazards model applied to the pooled data as the benchmark, both ODAC and the proposed federated pseudo-value regression produce bias distributions centered close to zero. While the proposed approach exhibits slightly greater variability than ODAC—consistent with the fact that pseudo-value regression provides an approximation to the Cox partial likelihood estimator—the magnitude of bias remains negligible across different network configurations and event-rate scenarios. These findings support the accuracy and validity of the proposed method under the PH assumption. In contrast to ODAC, which requires participating sites to share the set of unique event times across the network to construct the surrogate partial likelihood, the proposed framework avoids transmitting such survival information.

\subsection{Scenario 2: Time-Varying Treatment Effects}
To evaluate model performance when the PH assumption is violated, we generated survival data from a Weibull model incorporating a binary treatment indicator $A$ and a continuous confounder $X$. The shape parameter was defined as a function of treatment, $\alpha(A) = \alpha_0 + \gamma A$, while the scale parameter followed $\lambda(X,A) = \lambda_0 \exp(\beta_X X + \beta_A A)$. This specification induces a time-dependent HR, thereby violating the PH assumption when $\gamma \neq 0$.

We simulated $N = 5{,}500$ subjects distributed across $K = 20$ sites, with site-specific sample sizes ranging from 50 to 3,000. For each subject, we simulated a continuous confounder $X \sim \mathcal{N}(0,1)$ and assigned a binary treatment $A$ via $\text{logit}\{P(A=1 \mid X)\} = 0.5X$. Parameters were set to $\alpha_0 = 0.8$, $\gamma = 0.5$, $\lambda_0 = 15$, $\beta_X = \log(0.7)$, and $\beta_A = \log(1.5)$.

\begin{figure}[htbp]
  \centering
  \includegraphics[width=0.9\textwidth]{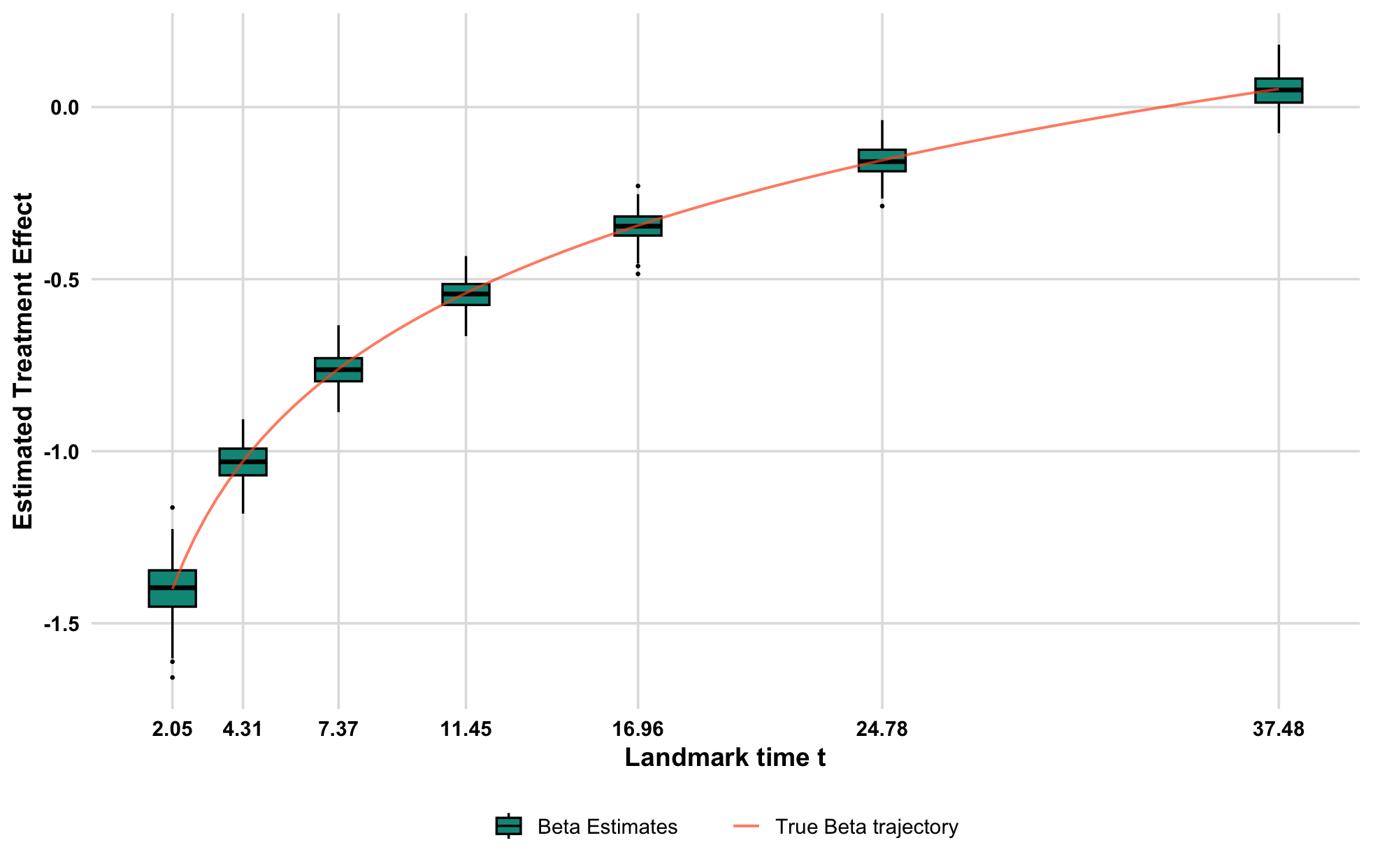} 
  \caption{Estimation of time-varying treatment effects under non-PH settings. The orange curve represents the true time-varying log-HR induced by the Weibull data-generating mechanism. Green boxplots show the distribution of estimates from the federated model at landmark times corresponding to the 20th--80th percentiles of the survival distribution ($N=5{,}500, K=20$).}
  \label{fig:sim_nph}
\end{figure}

Time-varying effects were estimated by constructing pseudo-observations at $J = 7$ landmark times and fitting a renewable GEE model with treatment-by-time interactions. Figure~\ref{fig:sim_nph} compares the estimated coefficients with the true log-HR trajectory. The federated estimates closely follow the analytical curve, demonstrating accurate recovery of time-varying treatment effects under non-PH.

\subsection{Scenario 3: Handling Sparse Site Heterogeneity}

In contrast to the preceding homogeneous settings, real-world federated networks often exhibit \emph{sparse site heterogeneity}, in which most sites share a common population-level covariate effect while a small subset displays systematic departures. To reflect this structure, we generated site-specific treatment effects using a sparse-mixture model:\begin{equation}\beta^{(k)}_{\mathrm{treat}} = \log(1.5) + \theta_k, \qquad \theta_k \sim (1-\pi)\delta_0 + \pi \mathcal{N}(0,\tau^2),\label{eq:sparse_mixture}\end{equation}where $\pi$ represents the proportion of heterogeneous sites and $\tau$ governs the magnitude of these local deviations. In this scenario, we fixed the total sample size at $N=5,000$ and the per-site sample size at $n_k = 100$, resulting in a stable network of $K = 50$ participating sites. We compared the global federated estimator, the local site-wise estimator, and the proposed debiased estimator in terms of their accuracy in recovering the true site-specific treatment effects.

\begin{figure}[htbp]
\centering
\includegraphics[width=0.9\textwidth]{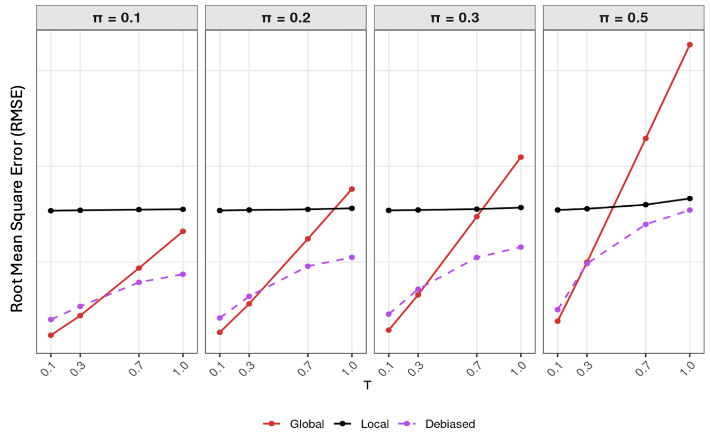}
\caption{Simulation-average RMSE of site-specific treatment effect estimates under sparse site heterogeneity ($K=50, n_k=100$). The horizontal axis shows the outlier magnitude $\tau$, and panels correspond to increasing proportions of heterogeneous sites $\pi$. We compare the global federated estimator (red line), the local site-wise estimator (black line), and the proposed debiased estimator with SURE-selected soft-thresholding (purple dashed line). Each point represents an average over 500 simulation replicates.}
\label{fig:debiasing}
\end{figure}

Under conditions of perfect homogeneity ($\pi = 0$), the global federated estimator achieves the highest precision by efficiently pooling all available data. In this baseline setting, the proposed debiased estimator demonstrates a minor loss in efficiency; this is attributed to the redundant denoising step inherent in the soft-thresholding procedure when no true site-level heterogeneity is present. However, as both the magnitude ($\tau$) and prevalence ($\pi$) of outlier sites increase, the global estimator's performance deteriorates sharply. While the local estimator remains stable, it suffers from high variance due to the limited sample size within each site. Figure~\ref{fig:debiasing} reports the simulation-average RMSE as a function of the heterogeneity magnitude $\tau$, stratified by the proportion of outlier sites $\pi$. 

Across these heterogeneous scenarios, the proposed debiased estimator consistently achieves the lowest RMSE by adaptively preserving genuine local deviations while shrinking noise-driven fluctuations toward the global effect. This robust bias–variance trade-off allow our method to significantly outperform both purely global and purely local estimation strategies as site-specific heterogeneity becomes more pronounced.

\section{Application to CAPriCORN Data}

The Chicago Area Patient-Centered Outcomes Research Network (CAPriCORN) \citep{kho2014} is a multi-institutional EHR data network in the Chicago area. As a real-world application, we analyzed data from the CAPriCORN pediatric cohort to investigate risk factors for the time to onset of childhood obesity, defined as achieving a body mass index (BMI) at or above the 95th percentile. The analysis included four participating hospitals with site-specific sample sizes of 24,305, 11,721, 6,696, and 3,143, respectively.

We treated federated pseudo-observations for the survival outcome (time to obesity) as outcomes in the renewable regression models. The analysis adjusted for baseline covariates including age, sex, race/ethnicity, BMI percentile, insurance type, comorbidity status, diastolic blood pressure, and systolic blood pressure. Table~1 summarizes these baseline characteristics by site.

\renewcommand{\arraystretch}{1.2}
\begin{table*}[!bp]
\begin{center}
\caption{Baseline Demographic and Clinical Characteristics by Site}
\label{tab:baseline}
\begin{tabular}{@{}lllll@{}}
\toprule
\textbf{Characteristic (Measure)} 
  & {\textbf{Site\,1}} 
  & {\textbf{Site\,2}} 
  & {\textbf{Site\,3}} 
  & {\textbf{Site\,4}} \\
\midrule  

\textbf{Number of subjects} (N)    
  & 24,305 & 11,721 &  6,696 & 3,143 \\
\textbf{Age} (Years, Mean)              
  & 5.2 & 5.9 & 5.4 & 6.4 \\
\textbf{BMI percentile} (Mean)          
  &   47.7 &   51.0 &   55.7 &   51.1 \\
\textbf{Diastolic BP} (mmHg, Mean)     
  &   59.4 &   60.2 &   58.8 &   60.4 \\
\textbf{Systolic BP} (mmHg, Mean)      
  &   98.1 &  101.3 &   97.7 &  100.5 \\
\addlinespace
\multicolumn{5}{@{}l}{\textbf{Gender}}\\
  \quad Female (N \%)          
    & 11,539 (47.5\%) & 5,558 (47.4\%)  & 3,202 (47.8\%) & 1,325 (42.2\%) \\
  \quad Male (N \%)            
    & 12,766 (52.5\%) & 6,163 (52.6\%)  & 3,494 (52.2\%) & 1,818 (57.8\%) \\
\addlinespace
\multicolumn{5}{@{}l}{\textbf{Race and Ethnicity}}\\
  \quad White (N \%)            
    &  1,941 (8.0\%) & 1,094 (9.3\%)  & 3,421 (51.1\%) & 1,178 (37.5\%) \\
  \quad Black (N \%)            
    & 16,999 (69.9\%) & 4,217 (36.0\%)  &  247 (3.7\%) &  702 (22.3\%) \\
  \quad Asian (N \%)            
    &  1,429 (5.9\%) & 5,428 (46.3\%)  & 2,776 (41.5\%) & 1,093 (34.8\%) \\
  \quad Hispanic (N \%)        
    &  1,330 (5.5\%) &  659 (5.6\%)  &  142 (2.1\%) &   81 (2.6\%) \\
  \quad Other (N \%)            
    &  2,606 (10.7\%) &  323 (2.8\%)  &  110 (1.6\%) &   89 (2.8\%) \\
\addlinespace
\multicolumn{5}{@{}l}{\textbf{Comorbidity}}\\
  \quad No (N \%)              
    & 23,633 (97.2\%) & 9,673 (82.5\%)  & 6,511 (97.2\%) & 2,751 (87.5\%) \\
  \quad Yes (N \%)             
    &   672 (2.8\%) & 2,048 (17.5\%)  &  185 (2.8\%) &  392 (12.5\%) \\
\addlinespace
\multicolumn{5}{@{}l}{\textbf{Obesity}}\\
  \quad No (N \%)              
    & 22,099 (90.9\%) & 9,826 (83.8\%)  & 3,994 (59.6\%) & 2,283 (72.6\%) \\
  \quad Yes (N \%)             
    &  2,206 (9.1\%) & 1,895 (16.2\%)  & 2,702 (40.4\%) &  860 (27.4\%) \\
\bottomrule
\end{tabular}
\end{center}
\end{table*}
\renewcommand{\arraystretch}{1.0}

Prior to model fitting, we assessed the proportional hazards (PH) assumption for each covariate using Schoenfeld residuals from site-specific Cox PH model. The diagnostics indicated violations of the PH assumption for baseline \textit{age} and \textit{BMI percentile} at one or more centers. To accommodate these departures from proportional hazards, we incorporated interaction terms between landmark time indicators and both \textit{age} and \textit{BMI percentile}, allowing time-varying coefficients for these variables while retaining time-invariant effects for the remaining covariates.


After accounting for these PH violations, we fit and compared pooled and federated pseudo-observation regression models. For the federated analysis, we first constructed a federated Kaplan–Meier estimator of the event-time distribution, from which we selected a grid of five landmark times ranging between 1.64 and 5.93 years. These points were chosen to fall within regions with an adequate number of patients at risk and stable survival estimates. At each time point, pseudo-observations were generated locally at each site and utilized as outcomes in a renewable generalized linear model to obtain the federated global regression estimates.

\begin{figure}[!bp]
  \centering
  \includegraphics[width=5.5in]{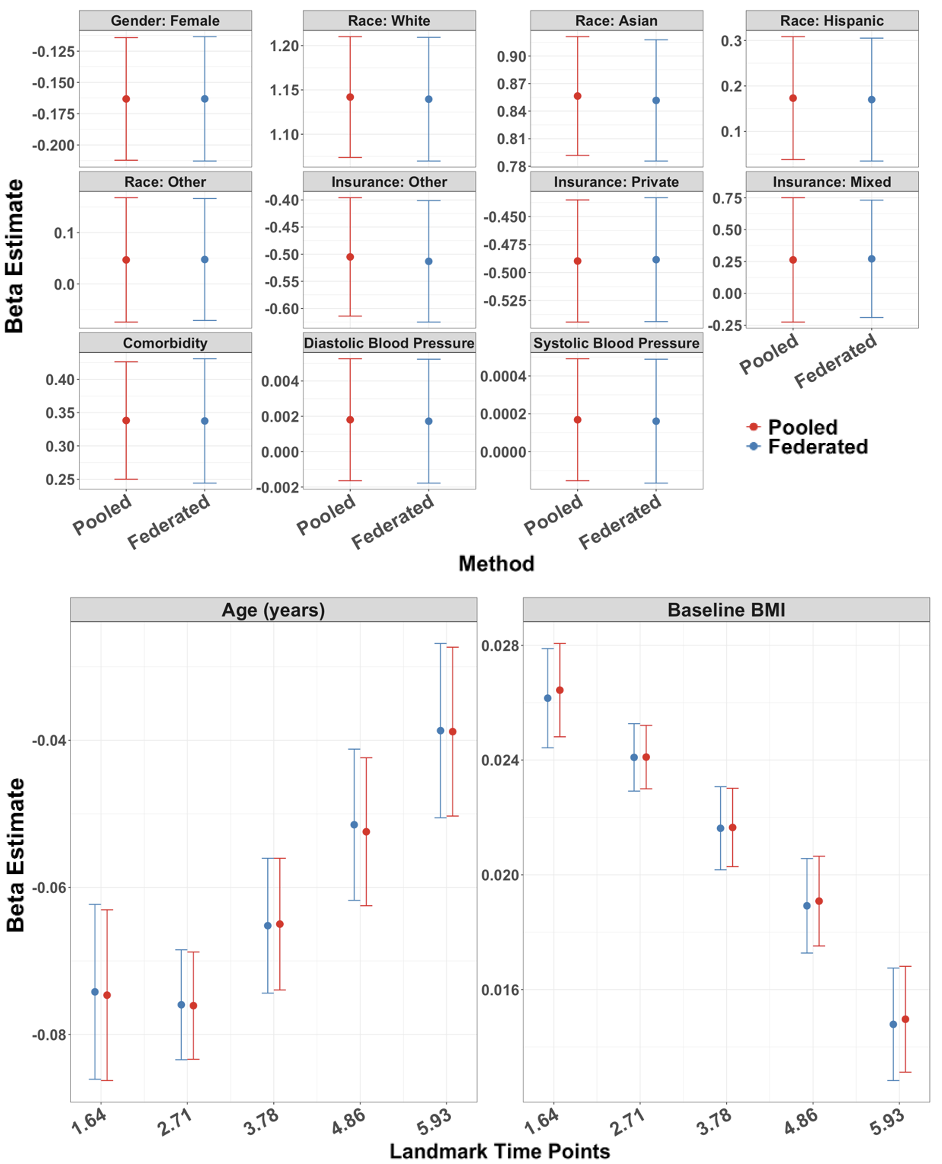}
  \caption{  
  Forest plots of point estimates with 95\% confidence intervals for each baseline covariate (excluding age and BMI percentile) from the pooled (red) and federated (blue) pseudo-observation regressions. Panels are faceted by covariate and illustrate that the federated algorithm recovers the same constant log-hazard ratios as the pooled analysis. Below, time-varying coefficient trajectories for baseline age and BMI percentile are shown over a grid of prespecified landmark times; each panel overlays federated (blue) and pooled (red) log-hazard ratio estimates, demonstrating near-identical temporal patterns.
  }
   \label{fig:forest_pooled_fed}
\end{figure}

Figure~\ref{fig:forest_pooled_fed} compares the coefficient estimates obtained from the centralized pooled Cox model and our proposed federated algorithm. The top panels present the time-invariant estimates ($\hat{\boldsymbol{\beta}}$) for baseline covariates, while the bottom panels display the trajectories of the time-varying coefficients for \textit{age} and \textit{BMI percentile} across five prespecified landmark times. Across all parameters, the federated estimates exhibit tight alignment with the pooled results, demonstrating that our distributed approach successfully replicates the centralized analysis. Clinically, the time-varying trajectories reveal that the coefficient for \textit{age} becomes less negative over time, whereas the effect of \textit{BMI percentile} decreases steadily, suggesting that baseline BMI plays a stronger role in near-term obesity risk.

\begin{figure}[!bp]
  \centering
  \includegraphics[width=5.5in]{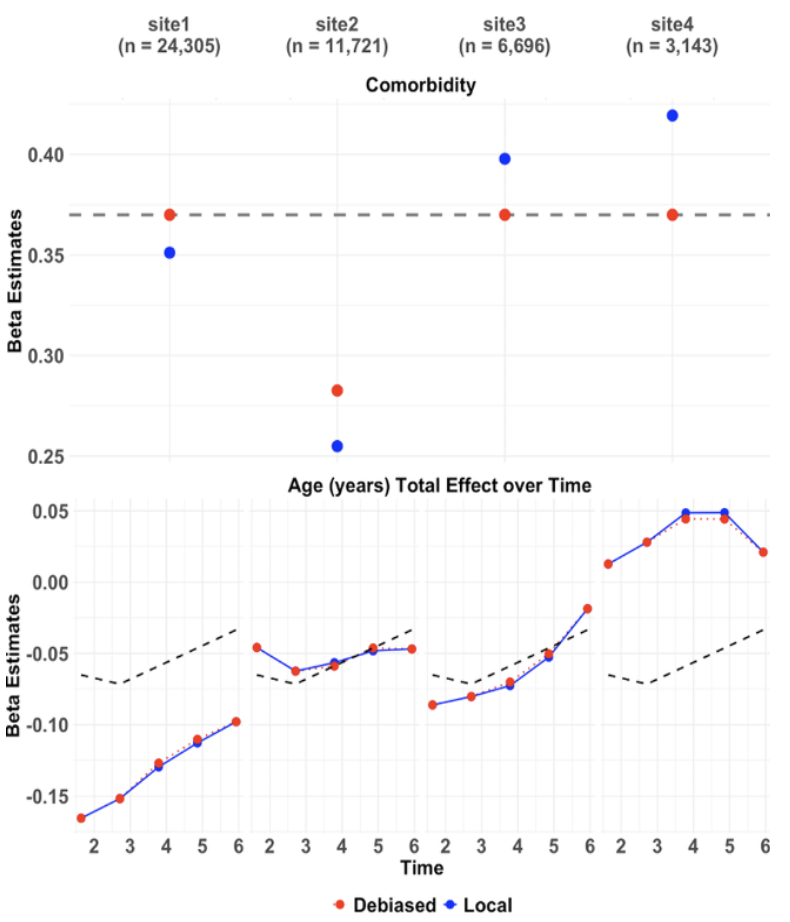}
  \caption{  
  Comparison of local (blue) and debiased (red) beta estimates across four sites for two covariates: comorbidity (top) and age (bottom). The dashed lines indicate the federated global estimates, which serve as the baseline reference for shrinkage. In the top panel, the vertical movement from a blue dot to a red dot visually represents the degree of shrinkage applied to the time-invariant comorbidity effect at each site. In the bottom panel, the trajectories plotted against follow-up time illustrate how the local time-varying effects of age are adjusted at each prespecified landmark time toward the global trajectory.
  }
  \label{fig:debiased_traj}
\end{figure}

Following estimation of the global federated model, we applied the variance-adaptive soft-thresholding procedure to obtain site-specific adjusted estimates. Figure~\ref{fig:debiased_traj} compares the original local, global, and adjusted estimates obtained via variance-adaptive soft-thresholding. For Comorbidity, local estimates on Sites 1, 3, and 4 are aggressively regularized to the global effect, whereas Site 2 retains its meaningful local signal. Conversely, the time-varying Age trajectories are largely preserved across all sites with minimal shrinkage, reflecting the high precision (low deviation variance) of these estimates supported by the full patient cohort. These results demonstrate the procedure's ability to filter noise while preserving genuine site-level heterogeneity

\section{Discussion and Conclusion}

We developed a federated pseudo-observation framework for multicenter time-to-event analysis when individual-level data cannot be centralized. The proposed method combines federated pseudo-observations from a distributed Kaplan--Meier estimator with renewable generalized linear models to enable communication-efficient regression modeling without sharing patient-level records. Unlike one-shot Cox-based federated approaches such as ODAC \citep{Duan2020}, it is not restricted to the proportional hazards assumption and avoids the exchange of sensitive survival information such as unique event times across sites. The framework accommodates both time-invariant and time-varying covariate effects, and valid statistical inference is supported by a robust sandwich variance estimator that accounts for the clustered dependence induced by repeated pseudo-observations.

Across simulation studies, the proposed method showed good operating characteristics in several settings relevant to multicenter survival analysis. Under proportional hazards, the federated pseudo-value regression produced estimates closely aligned with those from the pooled Cox model and ODAC, with comparable bias and variability across a range of event rates and site configurations. When the proportional hazards assumption was violated, the method accurately recovered time-varying log-hazard ratio trajectories using landmark-based pseudo-observations. In settings with sparse site heterogeneity, the proposed debiasing procedure improved estimation of site-specific effects by shrinking noise-driven local deviations toward the global estimate while preserving meaningful departures supported by the data. 

In the CAPriCORN pediatric obesity application, the proposed framework closely reproduced the corresponding pooled analysis while allowing greater modeling flexibility. In particular, violations of the proportional hazards assumption for baseline age and BMI percentile motivated the use of time-varying effects, and the federated pseudo-observation model recovered coefficient trajectories that were closely aligned with those from the pooled analysis. The heterogeneity-adjustment step further allowed site-specific deviations in covariate effects to be examined in a principled way, while limiting instability from noisy local estimates.

Several limitations should be noted. First, as with other pseudo-observation-based approaches, the method targets marginal survival functionals through an approximation rather than the Cox partial likelihood directly. Second, the pseudo-observation framework represents time-varying effects through a finite set of landmark times, so the estimated trajectories depend on the choice and number of evaluation points. Third, the proposed debiasing procedure is motivated by sparse heterogeneity and may be less suitable when heterogeneity is widespread or follows more complex dependence structures across sites. Moreover, while GSURE offers a principled data-driven rule for selecting the shrinkage parameter under correlated deviations, its performance may be less stable when the number of sites is small.

Overall, the proposed framework provides a flexible and privacy-preserving approach for multicenter survival analysis, while maintaining estimation and inference close to those from pooled analyses. These features make it useful in collaborative clinical research settings where direct sharing of individual-level time-to-event data is infeasible, but valid and interpretable survival modeling remains essential.

\section*{Acknowledgments}

The findings reported in this paper were enabled through a collaboration with the Chicago Area Patient-Centered Outcomes Research Network (CAPriCORN). CAPriCORN is a partnership between healthcare and research institutions that provides data through a federated harmonized common data model and works jointly with a Patient Community Advisory Committee, community-based organizations (CBOs), and non-profit organizations committed to enabling and delivering patient-centered clinical research and public health projects.
 
We acknowledge CAPriCORN's partners, the Chicago Area Institutional Review Board (CHAIRb), which serves as the central IRB of record for CAPriCORN-supported research, and the Medical Research Analytics and Informatics Alliance (MRAIA), which serves as the network’s honest data broker.

\bibliographystyle{agsm}
\bibliography{citations}

@article{sherman2016,
  title={Real-world evidence—what is it and what can it tell us},
  author={Sherman, Rachel E and Anderson, Steven A and Dal Pan, Gerald J and Gray, Gerry W and Gross, Thomas and Hunter, Nina L and LaVange, Lisa and Marinac-Dabic, Danica and Marks, Peter W and Robb, Melissa A and others},
  journal={N Engl J Med},
  volume={375},
  number={23},
  pages={2293--2297},
  year={2016}
}

@article{zhao2020deep,
  title={Deep neural networks for survival analysis using pseudo values},
  author={Zhao, Lili and Feng, Dai},
  journal={IEEE journal of biomedical and health informatics},
  volume={24},
  number={11},
  pages={3308--3314},
  year={2020},
  publisher={IEEE}
}

@article{zhao2020deepBioinformatics,
  title={Deep neural networks for predicting restricted mean survival times},
  author={Zhao, Lili},
  journal={Bioinformatics},
  volume={36},
  number={24},
  pages={5672--5677},
  year={2020},
  publisher={Oxford University Press}
}

@article{kho2014,
  title={CAPriCORN: Chicago area patient-centered outcomes research network},
  author={Kho, Abel N and Hynes, Denise M and Goel, Satyender and Solomonides, Anthony E and Price, Ron and Hota, Bala and Sims, Shannon A and Bahroos, Neil and Angulo, Francisco and Trick, William E and others},
  journal={Journal of the American Medical Informatics Association},
  volume={21},
  number={4},
  pages={607--611},
  year={2014},
  publisher={BMJ Publishing Group}
}

@article{lu2021,
  title={Multicenter privacy-preserving cox analysis based on homomorphic encryption},
  author={Lu, Yao and Tian, Yu and Zhou, Tianshu and Zhu, Shiqiang and Li, Jingsong},
  journal={IEEE Journal of Biomedical and Health Informatics},
  volume={25},
  number={9},
  pages={3310--3320},
  year={2021},
  publisher={IEEE}
}

@article{li2023,
  title={Distributed Cox proportional hazards regression using summary-level information},
  author={Li, Dongdong and Lu, Wenbin and Shu, Di and Toh, Sengwee and Wang, Rui},
  journal={Biostatistics},
  volume={24},
  number={3},
  pages={776--794},
  year={2023},
  publisher={Oxford University Press}
}

@article{bonomi2020,
  title={Protecting patient privacy in survival analyses},
  author={Bonomi, Luca and Jiang, Xiaoqian and Ohno-Machado, Lucila},
  journal={Journal of the American Medical Informatics Association},
  volume={27},
  number={3},
  pages={366--375},
  year={2020},
  publisher={Oxford University Press}
}

@article{froelicher2021,
  title={Truly privacy-preserving federated analytics for precision medicine with multiparty homomorphic encryption},
  author={Froelicher, David and Troncoso-Pastoriza, Juan R and Raisaro, Jean Louis and Cuendet, Michel A and Sousa, Joao Sa and Cho, Hyunghoon and Berger, Bonnie and Fellay, Jacques and Hubaux, Jean-Pierre},
  journal={Nature communications},
  volume={12},
  number={1},
  pages={5910},
  year={2021},
  publisher={Nature Publishing Group UK London}
}

@article{spath2022,
  title={Privacy-aware multi-institutional time-to-event studies},
  author={Sp{\"a}th, Julian and Matschinske, Julian and Kamanu, Frederick K and Murphy, Sabina A and Zolotareva, Olga and Bakhtiari, Mohammad and Antman, Elliott M and Loscalzo, Joseph and Brauneck, Alissa and Schmalhorst, Louisa and others},
  journal={PLOS Digital Health},
  volume={1},
  number={9},
  pages={e0000101},
  year={2022},
  publisher={Public Library of Science San Francisco, CA USA}
}

@incollection{vogelsang2020,
  title={A secure multi-party computation protocol for time-to-event analyses},
  author={Vogelsang, Lennart and Lehne, Moritz and Schoppmann, Phillipp and Prasser, Fabian and Thun, Sylvia and Scheuermann, Bj{\"o}rn and Schepers, Josef},
  booktitle={Digital Personalized Health and Medicine},
  pages={8--12},
  year={2020},
  publisher={IOS Press}
}

@article{concato2022,
  title={Real-world evidence—where are we now?},
  author={Concato, John and Corrigan-Curay, Jacqueline},
  journal={New England Journal of Medicine},
  volume={386},
  number={18},
  pages={1680--1682},
  year={2022},
  publisher={Mass Medical Soc}
}

@article{huang2025,
  title={Optimal multitask linear regression and contextual bandits under sparse heterogeneity},
  author={Huang, Xinmeng and Xu, Kan and Lee, Donghwan and Hassani, Hamed and Bastani, Hamsa and Dobriban, Edgar},
  journal={Journal of the American Statistical Association},
  pages={1--14},
  year={2025},
  publisher={Taylor \& Francis}
}

@article{izmailova2018,
  title={Wearable devices in clinical trials: hype and hypothesis},
  author={Izmailova, Elena S and Wagner, John A and Perakslis, Eric D},
  journal={Clinical Pharmacology \& Therapeutics},
  volume={104},
  number={1},
  pages={42--52},
  year={2018},
  publisher={Wiley Online Library}
}

@article{daniore2024,
  title={From wearable sensor data to digital biomarker development: ten lessons learned and a framework proposal},
  author={Daniore, Paola and Nittas, Vasileios and Haag, Christina and Bernard, J{\"u}rgen and Gonzenbach, Roman and von Wyl, Viktor},
  journal={NPJ digital medicine},
  volume={7},
  number={1},
  pages={161},
  year={2024},
  publisher={Nature Publishing Group UK London}
}

@article{baser2023,
  title={Use of open claims vs closed claims in health outcomes research},
  author={Baser, Onur and Samayoa, Gabriela and Yapar, Nehir and Baser, Erdem and Mete, Fatih},
  journal={Journal of Health Economics and Outcomes Research},
  volume={10},
  number={2},
  pages={44},
  year={2023}
}

@article{eldar2008generalized,
  title={Generalized SURE for exponential families: Applications to regularization},
  author={Eldar, Yonina C},
  journal={IEEE Transactions on Signal Processing},
  volume={57},
  number={2},
  pages={471--481},
  year={2008},
  publisher={IEEE}
}

@article{ahn2022,
  title={A brief introduction to research based on real-world evidence: considering the Korean National Health Insurance Service database},
  author={Ahn, Eun Kyoung},
  journal={Integrative Medicine Research},
  volume={11},
  number={2},
  pages={100797},
  year={2022},
  publisher={Elsevier}
}

@article{anyaso2023,
  title={Pseudo-value regression of clustered multistate current status data with informative cluster sizes},
  author={Anyaso-Samuel, Samuel and Bandyopadhyay, Dipankar and Datta, Somnath},
  journal={Statistical methods in medical research},
  volume={32},
  number={8},
  pages={1494--1510},
  year={2023},
  publisher={SAGE Publications Sage UK: London, England}
}

@article{donoho1994,
  title={Ideal spatial adaptation by wavelet shrinkage},
  author={Donoho, David L and Johnstone, Iain M},
  journal={biometrika},
  volume={81},
  number={3},
  pages={425--455},
  year={1994},
  publisher={Oxford University Press}
}

@article{duan2020,
  title={Learning from local to global: An efficient distributed algorithm for modeling time-to-event data},
  author={Duan, Rui and Luo, Chongliang and Schuemie, Martijn J and Tong, Jiayi and Liang, C Jason and Chang, Howard H and Boland, Mary Regina and Bian, Jiang and Xu, Hua and Holmes, John H and others},
  journal={Journal of the American Medical Informatics Association},
  volume={27},
  number={7},
  pages={1028--1036},
  year={2020},
  publisher={Oxford University Press}
}

@article{luo2022odach,
  title={ODACH: a one-shot distributed algorithm for Cox model with heterogeneous multi-center data},
  author={Luo, Chongliang and Duan, Rui and Naj, Adam C and Kranzler, Henry R and Bian, Jiang and Chen, Yong},
  journal={Scientific reports},
  volume={12},
  number={1},
  pages={6627},
  year={2022},
  publisher={Nature Publishing Group UK London}
}

@article{cox1972,
  title={Regression models and life-tables},
  author={Cox, David R},
  journal={Journal of the Royal Statistical Society: Series B (Methodological)},
  volume={34},
  number={2},
  pages={187--202},
  year={1972},
  publisher={Wiley Online Library}
}

@article{luo2020,
  title={Renewable estimation and incremental inference in generalized linear models with streaming data sets},
  author={Luo, Lan and Song, Peter X-K},
  journal={Journal of the Royal Statistical Society Series B: Statistical Methodology},
  volume={82},
  number={1},
  pages={69--97},
  year={2020},
  publisher={Oxford University Press}
}

@article{lu2015,
    title = {WebDISCO: a web service for distributed cox model learning without patient-level data sharing},
    author = {Lu, Chia-Lun and Wang, Shuang and Ji, Zhanglong and Wu, Yuan and Xiong, Li and Jiang, Xiaoqian and Ohno-Machado, Lucila},
    journal = {Journal of the American Medical Informatics Association},
    volume = {22},
    number = {6},
    pages = {1212-1219},
    year = {2015},
}

@article{malcolm2025,
    title = {Distributed Kaplan-Meier Analysis},
    author = {Risk, Malcolm and Zhao, Lili and Shi, Xu},
    journal = {},
    volume = {},
    number = {},
    pages = {},
    year = {2025},
}

@article{andersen2010,
  title={Pseudo-observations in survival analysis},
  author={Andersen, Per Kragh and Pohar Perme, Maja},
  journal={Statistical methods in medical research},
  volume={19},
  number={1},
  pages={71--99},
  year={2010},
  publisher={SAGE Publications Sage UK: London, England}
}

@article{xie2012sure,
  title={SURE estimates for a heteroscedastic hierarchical model},
  author={Xie, Xianchao and Kou, SC and Brown, Lawrence D},
  journal={Journal of the American Statistical Association},
  volume={107},
  number={500},
  pages={1465--1479},
  year={2012},
  publisher={Taylor \& Francis}
}
\end{document}